# An AI-Driven Multimodal Smart Home Platform for Continuous Monitoring and Assistance in Post-Stroke Motor Impairment

Chenyu Tang, Ruizhi Zhang, Shuo Gao, *Senior Member, IEEE*, Zihe Zhao, Zibo Zhang, Jiaqi Wang, Cong Li, Junliang Chen, Yanning Dai, Shengbo Wang, Ruoyu Juan, Qiaoying Li, Ruimou Xie, Xuhang Chen, Xinkai Zhou, Yunjia Xia, Jianan Chen, Fanghao Lu, Xin Li, Ningli Wang, Peter Smielewski, Yu Pan, Hubin Zhao, and Luigi G. Occhipinti, *Senior Member, IEEE*

***Abstract*—At-home rehabilitation for post-stroke patients presents significant challenges, as continuous, personalized care is often limited outside clinical settings. Moreover, the lack of integrated solutions capable of simultaneously monitoring motor recovery and providing intelligent assistance in home environments hampers rehabilitation outcomes. Here, we present a multimodal smart home platform designed for continuous, at-home rehabilitation of post-stroke patients, integrating wearable sensing, ambient monitoring, and adaptive automation. A plantar pressure insole equipped with a machine learning pipeline classifies users into motor recovery stages with up to 94% accuracy, enabling quantitative tracking of walking patterns during daily activities. An optional head-mounted eye-tracking module, together with ambient sensors such as cameras and microphones, supports seamless hands-free control of household devices with a 100% success rate and sub-second response time. These data streams are fused locally via a hierarchical Internet of Things (IoT) architecture, ensuring low latency and data privacy. An embedded large language model (LLM) agent, Auto-Care, continuously interprets multimodal data to provide real-time interventions—issuing personalized reminders, adjusting environmental conditions, and notifying caregivers. Implemented in a post-stroke context, this integrated smart home platform increased mean user satisfaction from 3.9 ± 0.8 in conventional home environments to 8.4 ± 0.6 with the full system (n = 20). Beyond stroke, the system offers a scalable, patient-centered framework with potential for long-term use in broader neurorehabilitation and aging-in-place applications.**

This work was supported in part by the National Natural Science Foundation of China under Grant 62171014 (S. Gao), in part by the Beihang Ganwei Project under Grant JKF-20240590 (S. Gao), in part by The Royal Society under Research Grant RGS/R2/222333 (H. Zhao), in part by the Engineering and Physical Sciences Research Council (EPSRC) under Grants 13171178 R00287 (H.Zhao) and EP/K03099X/1, EP/L016087/1, EP/W024284/1, EP/P027628/1 (L.G. Occhipinti), in part by the European Innovation Council (EIC) under the European Union's Horizon Europe research and innovation program Grant 101099093 (H. Zhao), in part by The British Council under Contract 45371261 (L.G. Occhipinti), and in part by Endoenergy Systems under Grant G119004 and Haleon under CAPE partnership Grant G110480 (C.Tang).These authors contributed equally: Chenyu Tang, Ruizhi Zhang, Shuo Gao (Corresponding authors: Shuo Gao; Hubin Zhao; Luigi G. Occhipinti).

Chenyu Tang is with the Hangzhou International Innovation Institute, Beihang University, Hangzhou, China and Department of Engineering, University of Cambridge, Cambridge, UK (e-mail: ct631@cam.ac.uk).

Ruizhi Zhang, Zihe Zhao, Jiaqi Wang, Cong Li, Junliang Chen and Shengbo Wang are with the School of Instrumentation and Optoelectronic Engineering, Beihang University, Beijing 100191, China (e-mail: zhangruizhi1996@buaa.edu.cn; by1917059@buaa.edu.cn; katrina0625@buaa.edu.cn; cong_li@buaa.edu.cn; junliang_chen@buaa.edu.cn; wangshengb@buaa.edu.cn).

Shuo Gao is with the Hangzhou International Innovation Institute, Beihang University, Hangzhou, China and School of Instrumentation and Optoelectronic Engineering, Beihang University, Beijing, China (shuo_gao@buaa.edu.cn).

Zibo Zhang, and Luigi G. Occhipinti are with the Department of Engineering, University of Cambridge, Cambridge, UK (email: zz534@cam.ac.uk; lgo23@cam.ac.uk).

Yanning Dai is with the AI Initiative, KAUST, Thuwal, Saudi Arabia (email: yanning.dai@kaust.edu.sa).

Ruoyu Juan is with the Beijing New Guoxin Software Evaluation Technology Co ltd, Beijing, China (email: juanry@sic.gov.cn).

Qiaoying Li is with the Stomatology Department, Shijiazhuang People's Hospital, Shijiazhuang, China (email: rmyylqy@163.com).

Ruimou Xie, Xin Li, and Yu Pan are with the Department of Rehabilitation Medicine, Beijing Tsinghua Changgung Hospital, Tsinghua University, Beijing, China (email: xrma03496@btch.edu.cn; lxa02016@btch.edu.cn; panyu@btch.edu.cn).

Xuhang Chen and Peter Smielewski are with the Department of Clinical Neurosciences, University of Cambridge, Cambridge, UK (email: xc369@cam.ac.uk; ps10011@cam.ac.uk).

Xinkai Zhou, Yunjia Xia, Jianan Chen, and Hubin Zhao are with the HUB of Intelligent Neuro-engineering (HUBIN), CREATe, Division of Surgery and Interventional Science, University College London, Stanmore, UK (email: xinkai.zhou.21@ucl.ac.uk; yunjia.xia.18@ucl.ac.uk; jianan.chen.22@ucl.ac.uk; hubin.zhao@ucl.ac.uk).

Fanghao Lu is with the Hangzhou International Innovation Institute, Beihang University, Hangzhou, China (email: lufanghao@buaa.edu.cn).

Ningli Wang is with the Beijing Tongren Hospital, Capital Medical University, Beijing, China (email: wningli@vip.163.com).

## I. INTRODUCTION

Stroke is the third leading cause of disability worldwide, affecting more than 101 million people [1, 2]. Survivors often experience motor impairments (60–80%), cognitive deficits (20 – 30%), and speech difficulties (30 – 50%), which significantly compromise their independence and quality of life [3, 4]. Post-stroke recovery is not only a prolonged process but also a resource-intensive one, imposing significant economic and caregiving burdens on families and healthcare systems — a challenge exacerbated by global aging [5]. For many patients, the home becomes a critical environment for rehabilitation, as opportunities for continuous and personalized the need for innovative solutions that can support patients in their daily lives, monitor health conditions effectively, and adapt to individual needs [7, 8].

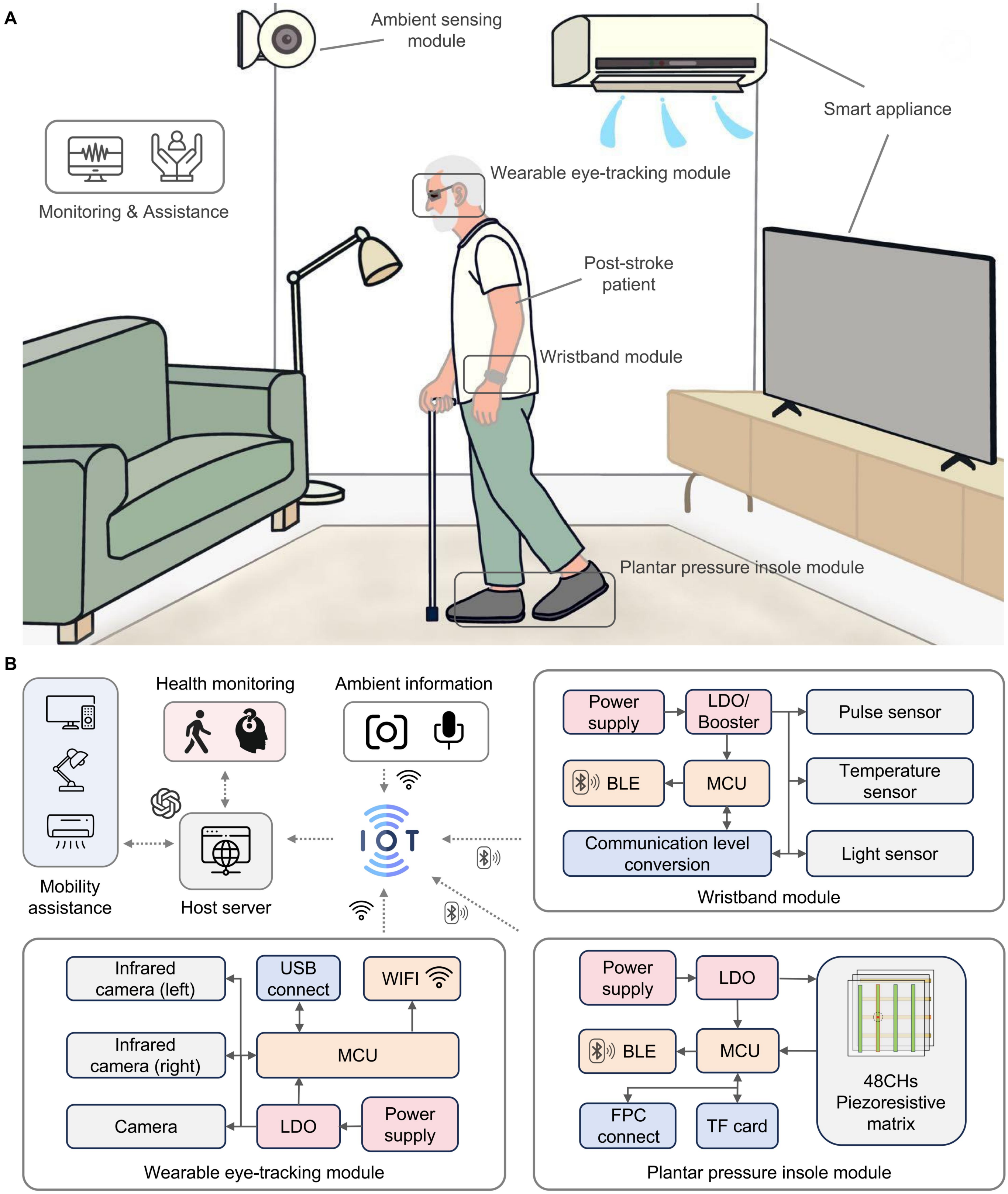

**Fig. 1.** Overview of the platform developed for at-home rehabilitation of post-stroke patients. A, Smart home setup. The system integrates wearable and ambient sensors, enabling real-time health monitoring and seamless interaction with smart appliances to support post-stroke recovery at home. B, System architecture and module design for multi-modal sensing. The platform comprises wearable modules for eye tracking, plantar pressure sensing, and physiological monitoring, along with a host server and ambient sensors to collect, process, and analyze multi-modal data for real-time monitoring and assistance.

TABLE I
CONCEPTUAL COMPARISON OF REPRESENTATIVE SMART HOME AND TELEREHABILITATION PLATFORMS.

| Study | Target population | Sensing modalities | Intelligent processing | Functionalities | Cohort | Integration level |
|---|---|---|---|---|---|---|
| [19] | Post-stroke patients (lower-limb rehabilitation) | 4 IMUs + 4 pressure insoles | Feature extraction + traditional ML classifier | Exercise type recognition and remote data upload | N/A | Partial (remote monitoring and feedback) |
| [17] | Stroke and traumatic brain injury survivors (upper-limb rehabilitation) | 6 IMUs | Traditional ML classifier | Quantitative assessment of upper-limb motor impairment and movement quality | N=37 | Partial (modular system for monitoring and clinical score estimation) |
| [27] | Post-stroke patients (lower-limb rehabilitation) | 3-axis load cells on force plate | Traditional ML classifier | Classification of motor impairment severity (severe vs. moderate) and detection of long-term recovery trends during sit-to-stand tasks | N=15 | Partial (single-modality evaluation system) |
| [25] | ALS patients (severe motor impairments) | Infrared eye-tracking glasses + RFID | Deep learning classifier | Eye-controlled smart home control | N=34 | Partial (smart home appliance control) |
| This work | Post-stroke patients (varying motor impairment levels) | Plantar pressure insoles + physiological wristband + Infrared eye-tracking glasses + ambient cameras and microphones | Deep learning classifier + LLM agent | Continuous motor recovery monitoring, hands-free smart home control, and autonomous assistance | N=20 | Fully integrated (real-time monitoring and assistance in home rehabilitation) |

Recent advancements in wearable sensors, internet of things (IoT), and artificial intelligence (AI) technologies have opened new possibilities for at-home health monitoring and assistance [9, 10, 11, 12]. Various wearable sensors — including force sensors [13, 14, 15, 16], accelerometers [17, 18, 19, 20], and eye trackers [21, 22, 23] — have shown promise in tracking motor recovery and capturing aspects of daily function such as gait stability, balance, and fine motor control for post-stroke rehabilitation. Complementing these monitoring tools, assistive technologies, including robotic aids and smart home systems, have been developed to address specific rehabilitation and daily living needs [24, 25, 26]. However, most existing approaches remain fragmented, targeting either narrow monitoring tasks or isolated assistance functions [27, 28, 29]. Few solutions offer an integrated framework that can continuously and intelligently monitor motor impairments in natural home environments while also delivering adaptive assistance, such as context-aware interaction and personalized interventions tailored to patient-specific needs and environmental conditions.

Here, we report a smart home system designed to enable continuous at-home rehabilitation of post-stroke patients with motor impairments, integrating health monitoring and assistive functionalities into a single platform (Fig. 1). In contrast to many existing solutions that require adherence to specialized tasks or laboratory-based protocols, our approach continuously and unobtrusively captures patient data through natural daily activities — such as walking or interacting with household objects—reflecting real-world usage while minimizing patient burden. By leveraging multi-sensor fusion, the system comprehensively addresses the diverse needs of patients with post-stroke motor impairments. For motor rehabilitation monitoring, our plantar pressure array, coupled with a machine learning model, evaluates motor recovery, achieving a classification accuracy of 94.1% across three rehabilitation states. Ambient sensors, including cameras and microphones, enable seamless and precise smart home control with an operational success rate of 100% and a latency below one second. This multimodal collaborative design ensures accessibility for a wide range of users, allowing them to select the most suitable interaction modality based on their physical abilities. Additionally, we introduce an autonomous assistive agent, Auto-Care, powered by a large language model (LLM), which analyzes multimodal data to provide timely interventions such as health reminders, environmental adjustments, or caregiver notifications, increasing overall user satisfaction by an average of 29% ($p < 0.01$) compared to scenarios without the agent. The developed IoT framework is also compatible with the integration of future functionalities, such as robotics modules to assist with hand rehabilitation. This system provides the first fully integrated framework for simultaneous health monitoring and intelligent assistance in post-stroke home rehabilitation (Table I), offering a pathway toward comprehensive, patient-centered management. In the future, it holds potential for broader applications in other chronic conditions, such as amyotrophic lateral sclerosis (ALS), Parkinson’s disease, and aging populations.

## II. Methods

### A. Participants

We recruited 20 post-stroke patients (14 male, 6 female; mean age 51.4 ± 9.8 years) under ethical approval (Committee for Medical Research Ethics, First Hospital of Shijiazhuang City, project number 2020036). Clinical characteristics are summarized in Table II. All participants provided informed consent. Patients performed natural walking trials on a flat surface while wearing the plantar pressure insoles. Motor recovery was clinically evaluated using the Fugl-Meyer Assessment (FMA) scale [30].

TABLE II
Subject Demographic Details

| ID | Age | Sex | BMI | FMA Score | Symptoms |
|---|---|---|---|---|---|
| **1** | 68 | Male | 22.1 | 91 | Minor left-sided hemiparesis |
| **2** | 45 | Male | 23.6 | 93 | Difficulty with fine motor skills |
| **3** | 56 | Male | 29.2 | 90 | Mild gait instability |
| **4** | 53 | Female | 22.6 | 88 | Occasional imbalance while walking |
| **5** | 51 | Female | 24.3 | 87 | Left knee valgus during gait |
| **6** | 45 | Male | 35.0 | 91 | Subtle hand tremors |
| **7** | 31 | Female | 20.4 | 90 | Slightly asymmetric gait |
| **8** | 32 | Male | 25.2 | 59 | Moderate left-sided weakness |
| **9** | 52 | Male | 24.5 | 75 | Gait asymmetry with compensatory mechanisms |
| **10** | 69 | Male | 21.6 | 78 | Moderate foot inversion on the right |
| **11** | 49 | Male | 24.5 | 71 | Difficulty with midline balance |
| **12** | 61 | Male | 19.8 | 51 | Notable knee hyperextension on standing |
| **13** | 56 | Female | 21.7 | 66 | Partial hand movement loss |
| **14** | 38 | Female | 20.7 | 63 | Frequent toe dragging while walking |
| **15** | 58 | Male | 23.4 | 31 | Severe left-sided hemiplegia |
| **16** | 39 | Male | 22.9 | 38 | Foot drop requiring ankle support |
| **17** | 36 | Male | 25.6 | 41 | Loss of balance with need for assistance |
| **18** | 68 | Male | 23.7 | 38 | Prolonged stance phase on unaffected side |
| **19** | 63 | Female | 21.8 | 36 | Severe spasticity in right arm and leg |
| **20** | 56 | Male | 26.1 | 30 | Complete paralysis of left side |

### B. Wearable devices

We developed three custom wearable devices, largely assembled from commercial sensing and processing modules, to enable continuous, multimodal monitoring in daily living environments.

*1) Plantar pressure insole:*

The insole integrates a 4 × 12 resistive pressure sensor array (48 sensors, density 0.23 sensors/cm²) laminated in a flexible multilayer structure. The design is identical to our earlier work [12, 29] and therefore only key details are summarized here. The array, approximately 100 $\mu$m thick, allows comfortable long-term wear. Data were sampled at 200 Hz, digitized by an HC32F460 microcontroller (ARM Cortex-M4), and transmitted wirelessly via a CH9141 Bluetooth module.

*2) Wristband module:*

The wristband was designed to collect auxiliary physiological and environmental data and is based on a modular assembly of commercial sensors. It integrates: (i) MAX30101 (Maxim Integrated) for photoplethysmography (PPG), (ii) AS7341 (AMS) for ambient light spectrum measurement, and (iii) MTS4B (MEMSIC) for skin temperature. An STM32L412 microcontroller (STMicroelectronics) coordinates sampling and communication, while data are transmitted through a CH9141 BLE module. The system is powered by a rechargeable lithium battery and supports more than 12 hours of continuous operation. The compact six-layer PCB (40 × 30 mm) and lightweight enclosure ensure wearability and robustness.

*3) Wearable eye tracker:*

The eye tracker provides optional assessment of visual attention and is also constructed from commercial modules. It employs two near-infrared cameras (Sony IMX258 sensors with integrated IR LEDs) for pupil and eye-corner tracking and one forward-facing IMX258 camera for scene capture. Processing is performed by an OrangePi CM5 module (RK3588S SoC, Rockchip), which computes gaze coordinates in real time using calibration data. Wireless data transmission is enabled by a CDW-20U5622 WiFi module. The device is powered by a rechargeable lithium battery, maintaining portability while achieving high-resolution (12 MP at 30 FPS) gaze tracking synchronized with scene video.

Together, these devices provide a lightweight and unobtrusive multimodal sensing platform that integrates plantar dynamics, auxiliary physiological parameters, and optional eye-tracking signals for home-based rehabilitation monitoring.

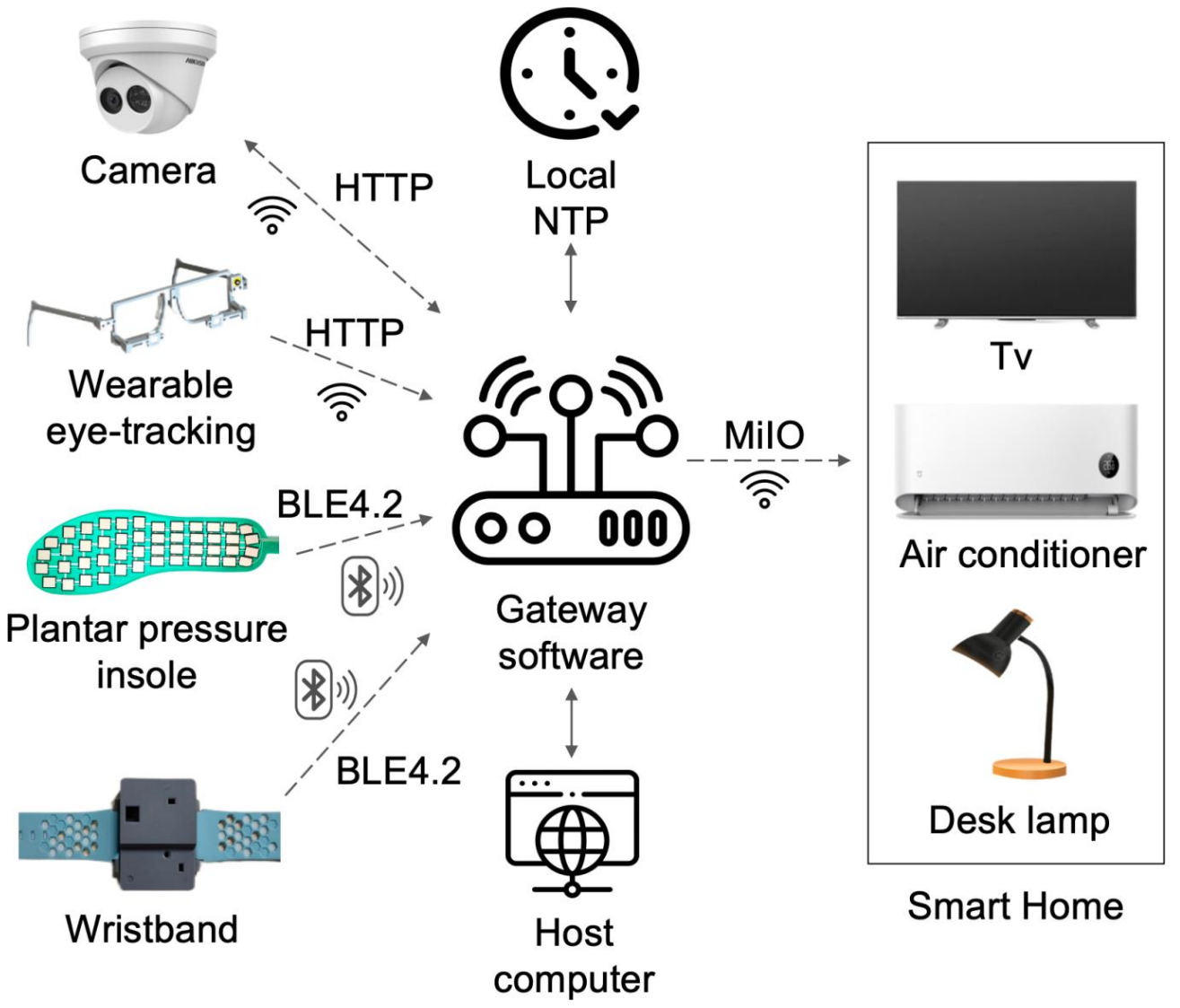

**Fig. 2.** Diagram of the hierarchical IoT architecture.

## C. IoT framework

To integrate heterogeneous sensing devices and enable real-time monitoring in home environments, we designed a hierarchical IoT framework (Fig. 2). The plantar pressure insoles and wristbands communicate with the gateway via Bluetooth Low Energy (BLE 4.2), while the wearable eye-tracking device and ambient camera transmit data through WiFi using HTTP. A custom gateway software running on a host computer aggregates multimodal data streams, synchronizes them via a local network time protocol (NTP) server, and performs lightweight fusion and preprocessing. Processed information is then distributed to smart home endpoints, including televisions, lamps, and air conditioners, via the MiIO protocol, enabling context-aware interaction. This layered architecture ensures robust, synchronized, and scalable communication among all modules, supporting continuous rehabilitation monitoring in real-world home settings. Processed multimodal data are further streamed into the Auto-Care module for high-level decision-making and intervention management.

## D. Data collection protocol and edge case handling

Walking sessions were carried out in home-like environments under medical supervision. Plantar pressure signals were continuously recorded at 200 Hz and segmented into 5-s windows for analysis. Each segment was labeled according to the participant's FMA score, and the dataset was constructed to maintain a balanced representation across mild, moderate, and severe impairment levels.

During deployment, three types of edge cases were frequently encountered:

1) Very short walks (fewer than three strides),

2) Assisted walking with canes or furniture support, and

3) Abrupt speed changes or multitasking during walking.

The first two cases were excluded through automated stride-length thresholds and auxiliary video verification, while the third was retained to improve model robustness.

## E. Motor impairment model design and training

As illustrated in Fig. 4C, the proposed motor impairment classification model adopts a dual-stream convolutional–fully connected architecture to capture both temporal dynamics and inter-limb asymmetries in plantar pressure patterns. A total of 1,543 gait segments were obtained across 20 participants. Each segment, sampled at 200 Hz for 5 s per foot (48 channels), was labeled according to the participant's Fugl–Meyer Assessment (FMA) score and categorized into three levels of motor impairment: mild (FMA $\geq$ 85), moderate (50 $\leq$ FMA $<$ 85), and severe (FMA $<$ 50). The dataset was randomly divided into 80% training and 20% testing subsets.

Raw plantar pressure signals from the left and right insoles were converted into 2D pressure maps (224 × 224 pixels) and separately encoded using two ResNet-101 convolutional backbones pretrained on ImageNet [43]. The encoder outputs (2048-dimensional feature vectors from each foot) were concatenated and passed into a multi-layer perceptron (MLP) classifier comprising the following fully connected layers [4096 → 1024 → 256 → 3], each followed by Batch Normalization [44], ReLU activation, and Dropout (p = 0.3) except the output layer, which uses softmax for 3-class prediction.

Model training was implemented in PyTorch 2.0, with cross-entropy loss as the objective function and the Adam optimizer (learning rate = $1 \times 10^{-4}$, weight decay = $1 \times 10^{-5}$). The batch size was 32, and training was performed on an NVIDIA RTX 4090 GPU. Early stopping (patience = 15 epochs) was applied based on validation loss to prevent overfitting. Training converged after approximately 80 epochs. The trained model outputs the predicted impairment level for each input segment, providing a continuous assessment of motor recovery from insole signals.

## F. Smart home control integration

To support real-time interaction between users and household appliances, the platform integrates smart home control via the MiIO protocol, enabling communication with commercial IoT devices such as lights, televisions, and air conditioners. The central gateway receives high-level control commands — derived from multimodal inputs such as voice, gaze, and ambient context — and translates them into actionable MiIO API requests. Each device is assigned a unique ID and mapped to a structured control schema (e.g., "toggle_light"), ensuring reliable and consistent operation. Device states are monitored and updated periodically to reflect actual execution and maintain synchronization with user intent. All data processing related to camera and microphone input is performed locally to preserve user privacy. Additionally, safety-critical actions are governed by fallback rule-based logic to prevent erroneous or ambiguous system behavior, such as multiple triggers or misinterpreted commands.

### *G. LLM agent integration*

To enable autonomous and context-aware assistance, the Auto-Care agent was implemented using the GPT-4o Mini API hosted on a local server. Multimodal sensor data — including physiological signals (e.g., heart rate, HRV, skin temperature), ambient conditions (e.g., light level, time of day), and user activity states (e.g., walking, sitting, falling) — were downsampled to 1-minute intervals and structured into standardized JSON prompts. Each prompt followed a fixed schema to enhance interpretability and ensure low-latency inference. The API was queried with a temperature setting of 0.7, selected to balance response diversity and consistency, particularly for scenario-driven intervention prompts. The agent processed each prompt using chain-of-thought reasoning, optionally guided by predefined intervention demos for recurring cases. Agent outputs, returned in structured JSON format, were parsed by the gateway software and translated into actionable smart home commands via the MiIO interface. To ensure reliability, all Auto-Care outputs were screened through a safety check layer before execution. Each generated command was first validated using predefined rules to reject incomplete or out-of-scope instructions, and then semantically verified against a whitelist of safe smart-home actions. In testing with 100 representative prompts, this safety layer detected all 7 erroneous outputs without triggering any false activations, confirming robust filtering performance.

## III. Results

### *A. Gait signal characterization across rehabilitation levels*

To evaluate the platform's capability for continuous monitoring of post-stroke motor recovery, 20 patients with different impairment levels were recruited and stratified into mild, moderate, and severe groups based on their FMA scores. During natural walking trials, the plantar pressure insoles captured gait dynamics with high spatial and temporal resolution.

Representative signals from each group are shown in Fig. 3.

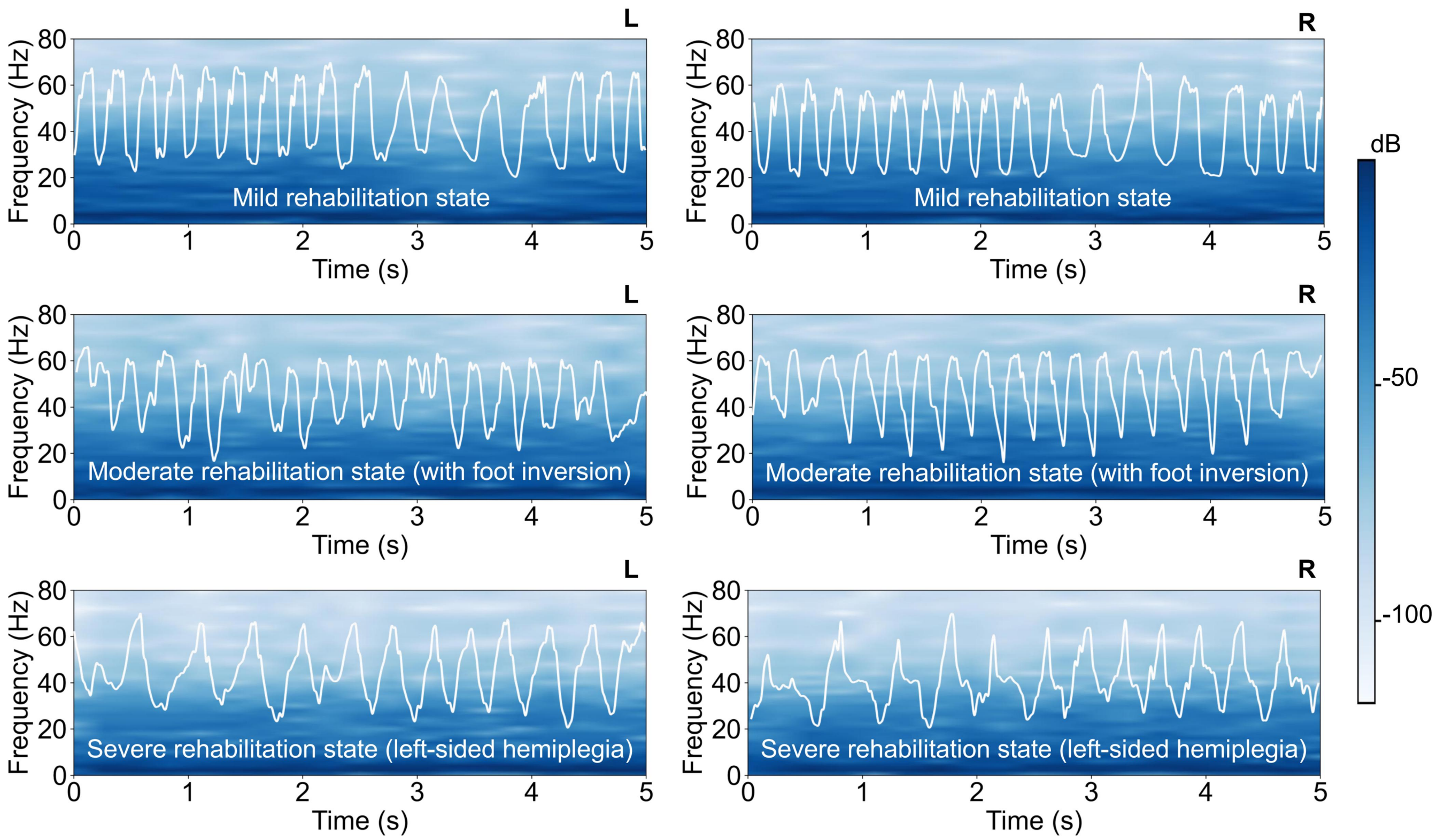

**Fig. 3.** Time-frequency spectrums of plantar pressure signals for left (L) and right (R) feet across mild, moderate, and severe rehabilitation levels.

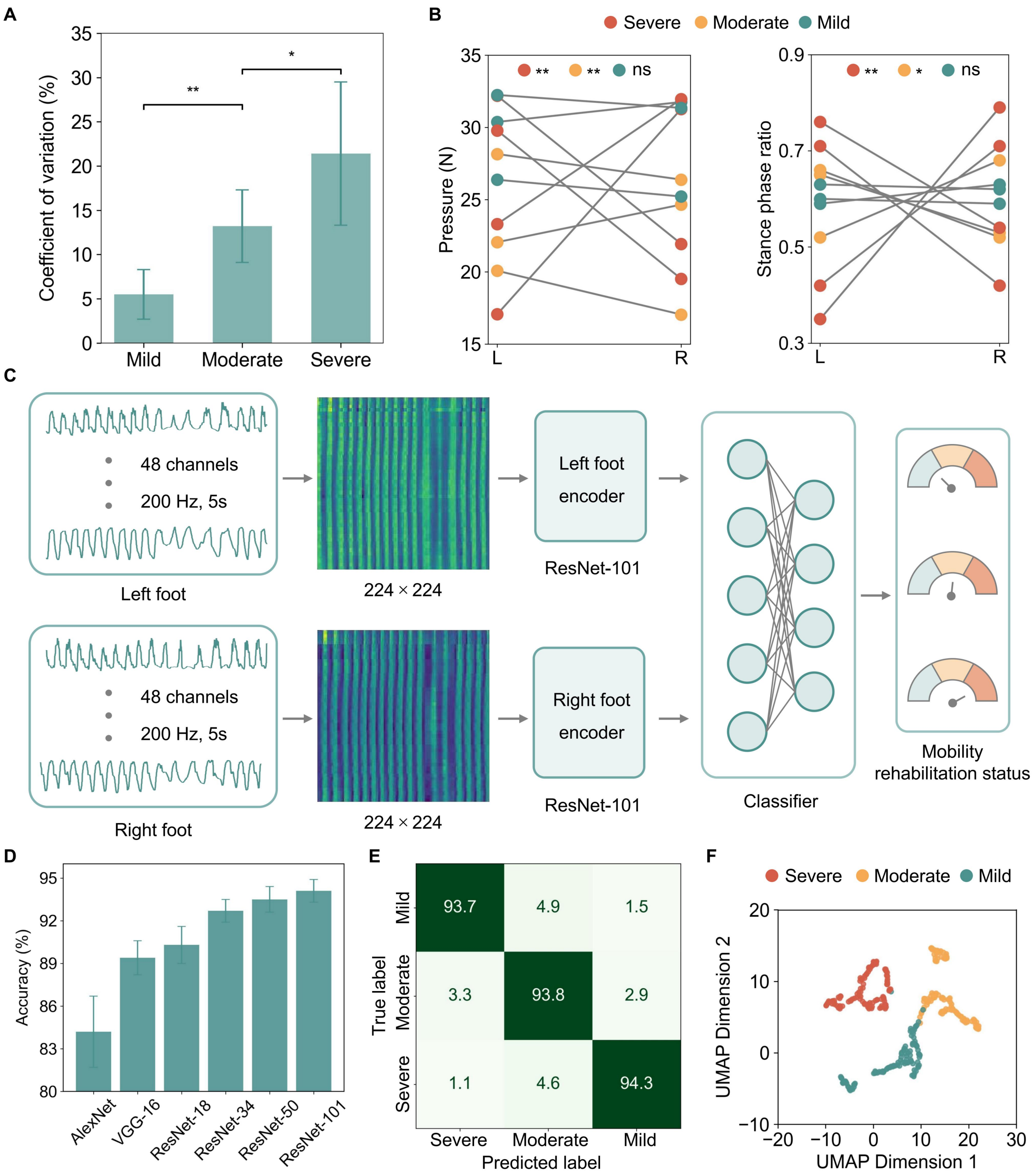

**Fig. 4.** Motor rehabilitation monitoring framework and performance evaluation. A, Coefficient of variation across rehabilitation stages. The variability in plantar pressure measurements increases with rehabilitation severity, highlighting distinct patterns among mild, moderate, and severe patients. B, Plantar pressure and stance phase ratio asymmetry. Left and right foot comparisons of pressure and stance phase ratio across rehabilitation stages, showing asymmetrical trends associated with severity. C, Deep learning framework for motor rehabilitation status classification. D, Classification accuracy of various deep learning models. E, Confusion matrix of classification results. F, UMAP visualization of latent features.

Patients in the mild group exhibited well-formed, symmetrical patterns resembling those of healthy individuals, indicating effective weight distribution and propulsion. Patients in the moderate group showed irregular oscillations and reduced symmetry, reflecting instability in stance and swing phases. Patients in the severe group demonstrated diminished signals on the affected side and exaggerated compensatory responses from the unaffected side, such as asymmetric loading or prolonged stance [29, 31, 32]. These distinct signal patterns provide objective, sensor-based markers of motor recovery stages.

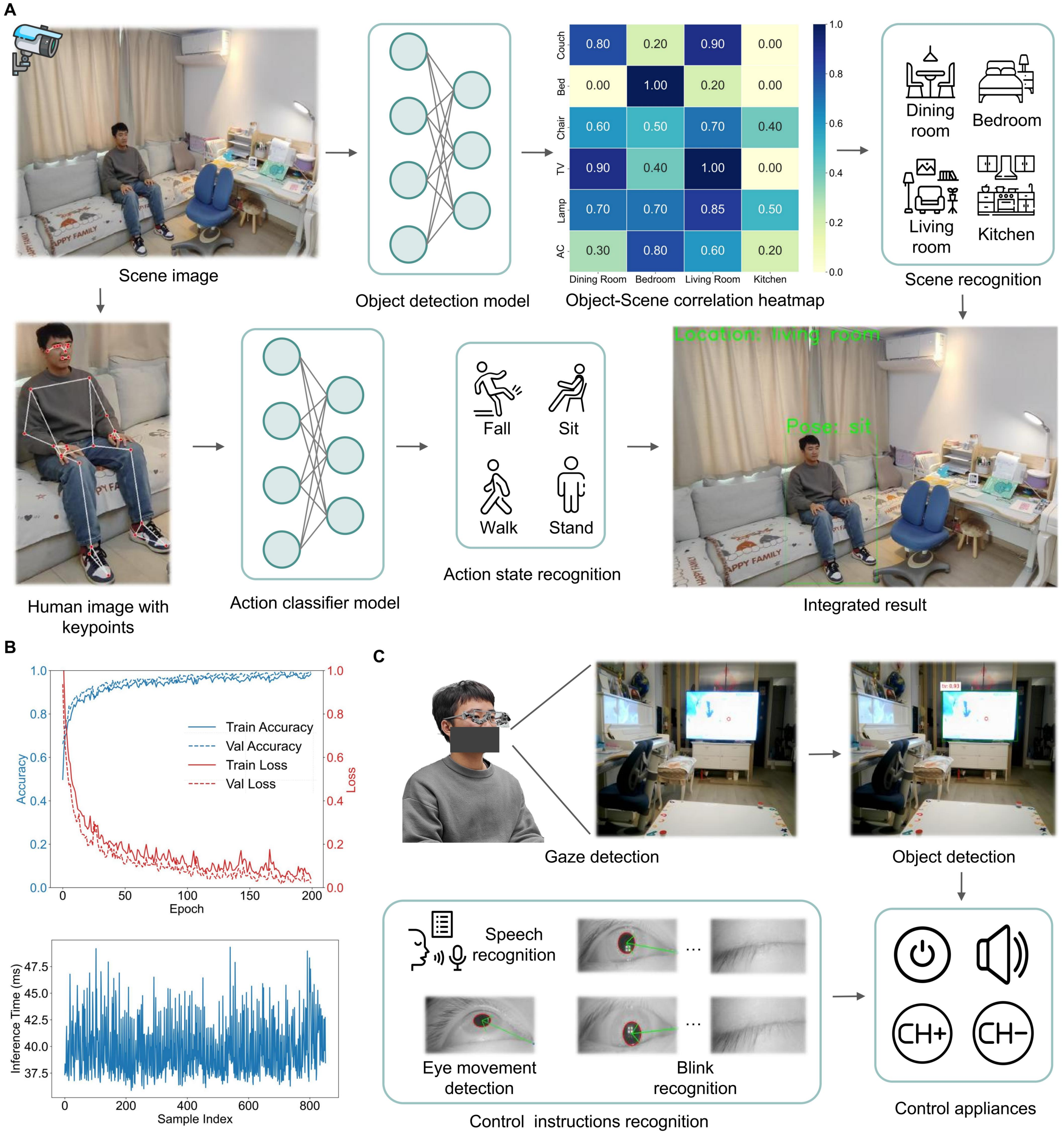

**Fig. 5.** Real-time scene detection, action recognition, and multimodal interaction system for stroke patient monitoring and smart home control. A, Scene images are analyzed using a fine-tuned YOLOv8n model for object detection and scene recognition, while human actions (e.g., sitting, walking, falling) are identified using a MediaPipe-based MLP classifier with 99.3% accuracy and <50 ms latency. Integrated results provide real-time feedback. B, Training curves and inference time confirm model accuracy and efficiency. C, Multimodal interaction combines gaze detection, blink recognition, and speech inputs to generate control commands for smart home devices, supporting diverse patient needs and enabling adaptive rehabilitation.

### *B. Continuous monitoring and classification performance*

Over a two-month observation period, we collected a total of 1,543 gait segments across all participants. At baseline, 6 patients were classified as severe, 7 as moderate, and 7 as mild. Longitudinally, one patient improved from severe to moderate and two patients improved from moderate to mild, confirming that the system can capture natural recovery trajectories over time.

To quantify gait characteristics, statistical features were extracted. Fig. 4A shows the coefficient of variation (CV) of plantar pressure, where severe patients exhibited significantly higher CV than moderate or mild patients, consistent with unstable gait. Fig. 4B demonstrates the asymmetry in stance phase ratio and pressure distribution between the two feet, with greater asymmetry observed in severe cases. These results confirm the platform's sensitivity to functional differences across recovery states.

The processing pipeline for automated state classification is outlined in Fig. 4C. Raw plantar pressure signals (5s segments) from the 48-channel insole arrays were converted into two-dimensional heatmaps (224 × 224 pixels) and processed by a ResNet-101 encoder. Outputs from the left and right foot encoders were fused in a MLP classifier to predict the rehabilitation state. This architecture effectively captures both temporal dynamics and inter-limb asymmetries, which are critical for motor recovery assessment.

Performance comparisons among candidate encoders are presented in Fig. 4D, where ResNet-101 achieved the highest accuracy and was selected as the optimal feature extractor. Quantitative evaluation based on the confusion matrix (Fig. 4E) shows that the optimized ResNet-101 + MLP architecture achieved an overall classification accuracy of 94.1%, computed using a weighted average reflecting the slightly larger number of mild segments in the dataset. The model also achieved a macro-averaged precision of 94.3%, recall of 93.9%, and F1-score of 94.1% across the three rehabilitation levels (mild, moderate, and severe). These results demonstrate balanced and robust performance across impairment categories, confirming the model's reliability for real-world rehabilitation monitoring. A UMAP visualization of encoded features (Fig. 4F) shows clear clustering of mild, moderate, and severe states, further confirming the discriminative power of the model. Together, these results highlight the system's ability to deliver accurate, continuous assessments of motor recovery during natural, unsupervised walking.

### *C. Smart home interaction via multi-sensor fusion*

Beyond monitoring, the platform was also designed to facilitate patient–environment interaction, enabling individuals with varying levels of speech and motor impairments to independently control household devices (a representative scenario is shown in Fig. 5A).

At its core, a fine-tuned YOLOv8n model processes live camera feeds to identify users and surrounding household objects with high accuracy [33], ensuring data privacy through localized inference. The system then infers contextual information about the current environment (e.g., living room, bedroom) by analyzing the spatial relationships among detected objects, as visualized in the object-scene correlation heatmap. Simultaneously, pose landmarks extracted via MediaPipe are transformed into normalized 3D coordinates and processed by an MLP-based action recognition model, enabling the system to classify user activity states (e.g., walking, sitting, falling) with 99.3% accuracy and a latency of <50 ms (Fig. 5B). These continuous activity logs are stored for retrospective analysis, contributing to long-term rehabilitation monitoring.

Building on this context-awareness, the system facilitates adaptive, multimodal interaction to accommodate a wide range of speech and motor impairments. Users with retained speech capabilities can issue voice commands via a microphone, which are processed by a lightweight ASR model (Whisper-tiny, 39M parameters [41]) and translated into device control instructions via the MiIO protocol. For individuals with severe speech impairments, a head-mounted eye tracker — equipped with infrared pupil and field-of-view cameras — computes real-time gaze coordinates after a brief calibration procedure. Gaze fixation across consecutive frames is then mapped onto detected household objects, enabling natural, hands-free control through eye movements with audio confirmation feedback (Fig. 5C).

These signals are synchronized within an IoT-based architecture that orchestrates sensor inputs and appliance responses in real time, thereby fostering user independence. By harmonizing scene detection, action recognition, and adaptive user interfaces, our approach enables a wide spectrum of post-stroke individuals to interact with their surroundings more seamlessly during daily rehabilitation.

### *D. LLM agent for autonomous assistance management*

To overcome the limitations of patients interacting with the platform solely based on subjective needs, we embedded an autonomous health management agent, Auto-Care, powered by GPT-4o Mini. By continuously analyzing multimodal data streams, the agent bridges the gap between passive monitoring and proactive intervention through context-aware decision-making. As shown in Fig. 6A, during gait training (point 1), the agent detected rising heart rate and temperature with decreasing HRV, recognized potential discomfort, and responded by pausing training, recommending hydration, and activating air conditioning. At point 2, when a fall occurred, the agent assessed the patient's condition via microphone input and, if necessary, alerted a caregiver. At point 3, the agent adapted smart lighting as ambient light decreased, ensuring sufficient illumination. These cases illustrate its ability to prioritize user safety, comfort, and continuity of daily activities.

Prompt optimization further enhanced the agent's performance (Fig. 6B, C). Chain-of-thought reasoning reduced decision ambiguity in complex scenarios [34], while pre-defined demos ensured consistency in recurring interventions. To maintain real-time responsiveness, a six-minute data context at 1-minute resolution was found to balance computational efficiency and inference precision.

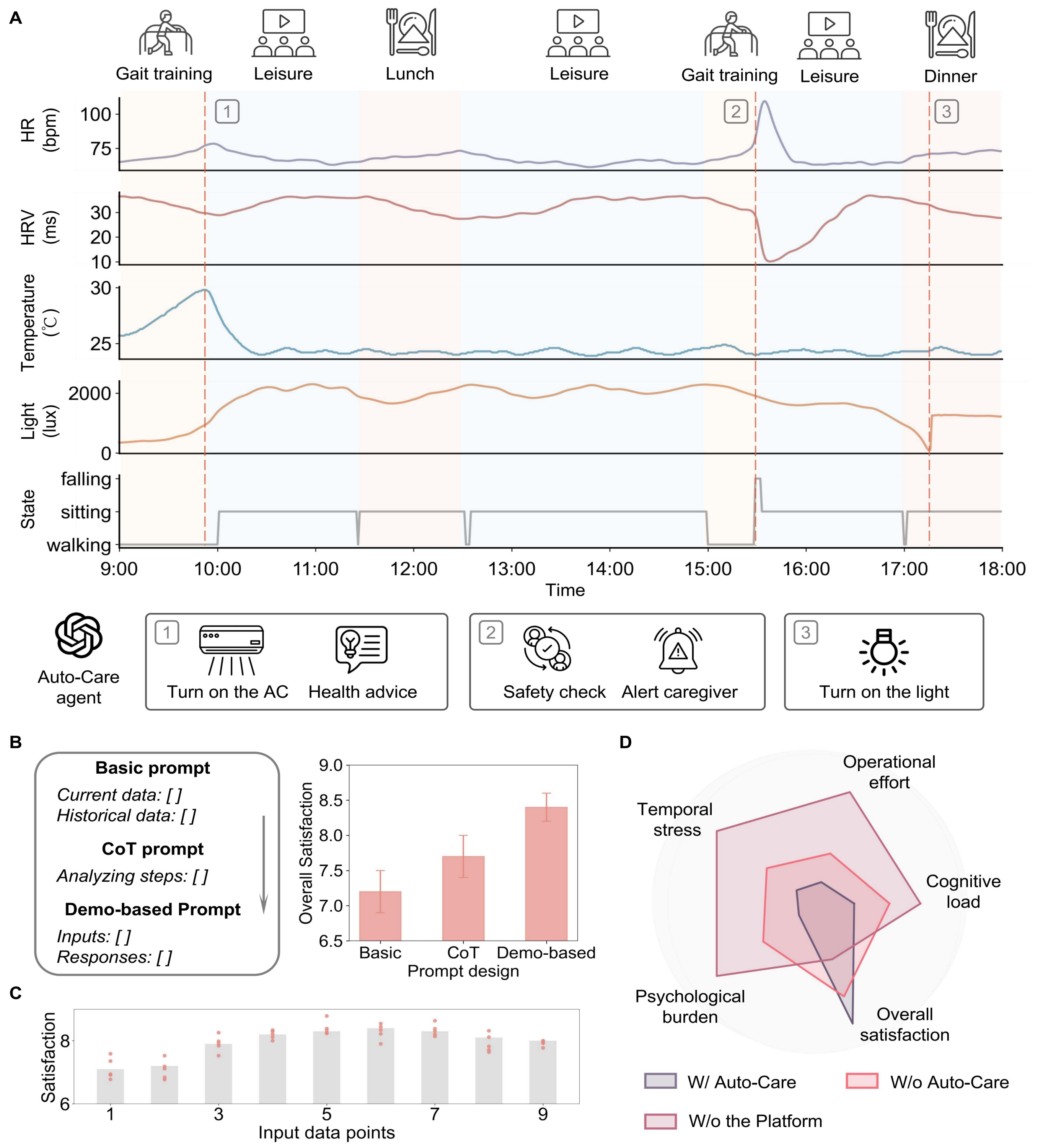

**Fig. 6.** LLM agent (Auto-Care) for autonomous assistance management. A, Daily monitoring and assistance by Auto-Care. Physiological and environmental signals, including heart rate (HR), heart rate variability (HRV), temperature, light intensity, and user state are continuously monitored. Auto-Care provides context-aware assistance such as safety checks, health advice, and environmental adjustments based on comprehensive analysis. B, Prompt design (basic, chain-of-thought (CoT), and CoT with demo-based prompts) and its impact on satisfaction. C, Effect of context length on agent performance. D, Radar plot comparing user satisfaction across various configurations (With Auto-Care, Without Auto-Care, and Without the platform).

TABLE III
EVALUATION CRITERIA FOR THE PLATFORM

| Criterion | Question | Scoring guidelines |
|---|---|---|
| **Cognitive load** | How much cognitive effort was required to complete the task? Was it mentally taxing? | 1 (Very low effort) to 10 (Very high effort) |
| **Operational effort** | How much physical or mental effort was needed to operate the system? Was it tedious or simple? | 1 (Very easy) to 10 (Very tedious) |
| **Temporal stress** | How much time pressure did you feel while completing the task? Was it urgent? | 1 (No pressure) to 10 (Extremely urgent) |
| **Psychological burden** | How comfortable or relaxed did you feel during the task? Were you anxious or calm? | 1 (Very anxious) to 10 (Very comfortable) |
| **Overall satisfaction** | How satisfied were you with the overall experience? Did the system meet your expectations? | 1 (Very dissatisfied) to 10 (Very satisfied) |

As summarized in Fig. 6D, Auto-Care significantly improved patient outcomes, with mean user satisfaction scores (10-point Likert scale) increasing from 3.9 ± 0.8 (95% CI: 3.53 – 4.27) in the baseline home environment to 6.5 ± 0.7 (95% CI: 6.17–6.83) with the platform without the agent, and further to 8.4 ± 0.6 (95% CI: 8.12 – 8.68) after integration of the Auto-Care agent (n = 20; Table III). These results demonstrate a substantial and statistically significant enhancement in perceived comfort, usability, and overall experience enabled by the intelligent assistance module. These findings highlight Auto-Care's potential to provide continuous, intelligent support tailored to patient needs, while balancing adaptability and efficiency in at-home rehabilitation.

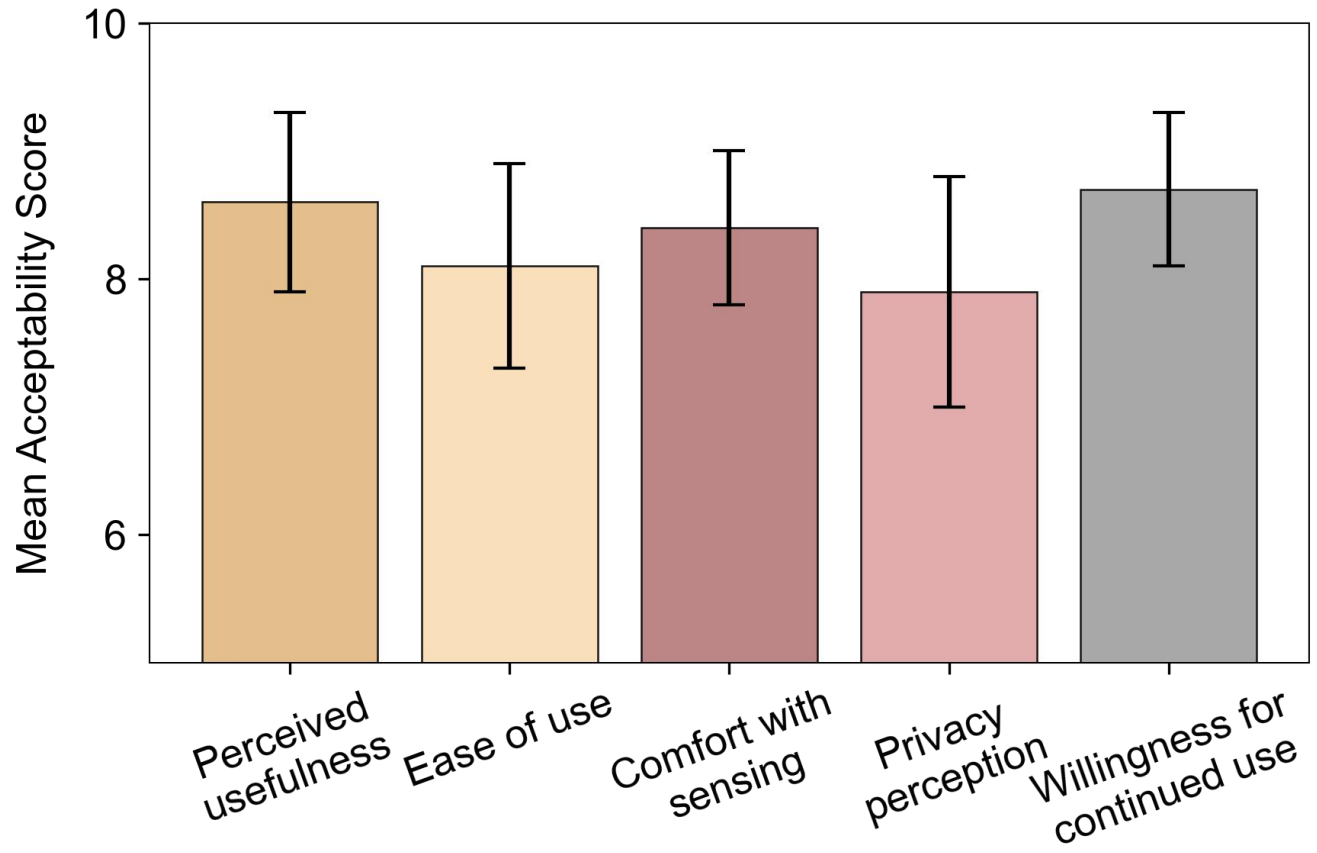

**Fig. 7.** User acceptability evaluation of the multimodal smart home platform (n = 20).

To assess user acceptability beyond satisfaction, participants completed a 10-point Likert-scale questionnaire evaluating perceived usefulness, ease of use, comfort, privacy perception, and overall willingness for long-term use (Fig. 7). Results indicate high overall acceptance (mean > 7 across all categories), particularly regarding usefulness and willing for continued use, while moderate concern was noted for privacy, which users found acceptable under local data-processing conditions.

## IV. DISCUSSION

This study presents a closed-loop, AI-driven smart home platform that enables continuous, personalized rehabilitation for post-stroke patients through integration of sensing, assessment, interaction, and autonomous assistance. Multimodal sensing — including plantar pressure insoles, wrist-based physiological monitoring, ambient cameras, and optional eye tracking — captures users' motor, physiological, and environmental states during daily activities. These data are processed in real time to classify motor impairment severity with 94.1% accuracy across mild, moderate, and severe recovery stages. Based on inferred state and context, the system enables hands-free smart home control with an average latency of 0.74 ± 0.18 s and a final success rate of 100% after possible command repetition. A locally deployed LLM-based agent, Auto-Care, interprets fused inputs to autonomously deliver interventions ranging from hydration reminders and environmental adjustments to caregiver alerts. Together, these modules form a sensing-understanding-action loop that bridges passive monitoring and proactive support, laying the foundation for scalable at-home neurorehabilitation.

The integration of multimodal sensing and intelligent automation represents a clear advance over existing methods. Traditional at-home rehabilitation relies on periodic visits and manual tracking, creating monitoring gaps and delayed interventions [35, 36]. In contrast, our platform continuously acquires data from wearables and ambient sensors to provide an uninterrupted, ecologically valid view of motor status, with data collection conducted alongside physicians using standardized scales to ensure reliability. Beyond improving clinical monitoring, the comprehensive multimodal dataset— spanning motor, physiological, and ambient signals — creates opportunities for discovery, such as novel biomarkers of recovery and longitudinal insights that inform individualized rehabilitation protocols and link home monitoring with clinical outcomes [37, 38]. In particular, this work follows the pathway outlined in our previously proposed framework for human body digital twins [9], which envisioned the integration of multimodal sensing and AI for continuous and personalized health management. The present study takes an initial step toward this vision by demonstrating real-time multimodal monitoring and adaptive assistance for post-stroke rehabilitation in natural home environments.

Despite these advances, several limitations remain. First, the present evaluation was conducted on a relatively small cohort of 20 post-stroke patients, constrained by the recruitment timeline and ethical requirements for in-home testing. Nevertheless, the dataset — comprising over 1,500 labeled gait segments across varying impairment levels — was sufficient to demonstrate the system's technical feasibility and the robustness of the proposed deep learning model. We acknowledge that a larger and more diverse cohort would further strengthen these findings, and additional patient recruitment is currently ongoing at our collaborating hospitals to support longitudinal and population-scale validation. While the current two-month deployment primarily establishes short-term feasibility, the system architecture and stable device performance indicate strong potential for sustained, long-term use in continuous home rehabilitation. Future studies should extend monitoring to capture long-term trajectories and develop predictive models of recovery. Although cognitive impairments are common post-stroke, our analysis emphasized motor status. Eye tracking offers potential for cognitive assessment but requires systematic validation. Future work could integrate cognitive measures, broadening the platform toward comprehensive management [39, 40]. Extending applicability to other chronic conditions, integrating robotic exoskeletons, and adopting edge computing will further improve efficiency, power use, and privacy.

In addition, data privacy and regulatory compliance are critical considerations for clinical translation. In the present proof-of-concept, the Auto-Care agent utilizes the GPT-4o Mini API with zero-retention enabled and HTTPS-encrypted communication, ensuring that no data are stored after inference and that transmissions are secure. Importantly, only de-identified contextual information is transmitted, while raw biosignals and personal identifiers remain locally processed. Although OpenAI services follow GDPR-aligned practices, they are not explicitly certified for HIPAA or GDPR healthcare compliance [42]. Besides, our system has been designed with modular model interfaces that can be substituted with on-device open-source LLMs (e.g., LLaMA3, DeepSeek) or enterprise-grade private deployments (e.g., Azure OpenAI with EU/UK data residency), enabling fully local inference where required. Such flexibility allows the platform to balance technical capability with regulatory safeguards, ensuring its adaptability to real-world healthcare settings.

In the long term, this platform may transform post-stroke care by supporting physical and psychological well-being. Continuous, personalized monitoring and intelligent assistance enable patients to regain independence, engage in social activities, and enhance life satisfaction. Its ability to analyze multimodal data also supports predictive analytics, enabling individualized recovery trajectories and risk assessment. By capturing subtle indicators — from motor to physiological and environmental factors — clinicians can design proactive interventions that refine rehabilitation strategies. Thus, the platform functions not only as a rehabilitation tool but also as a predictive health management system.